\documentclass[twocolumn]{aastex631}

\usepackage{xcolor}
\usepackage{amsmath}
\usepackage{hyperref}
\usepackage{xspace}
\usepackage{natbib}
\usepackage{graphicx}
\usepackage{tabularx, booktabs}

\newcolumntype{P}[1]{>{\centering\arraybackslash}p{#1}}

\graphicspath{{./figures/}{../figures/}{./}}

\newcommand{\lya}{Ly$\alpha$}
\newcommand{\AAGN}[1]{$A_\textnormal{AGN#1}$}

\shorttitle{The Impact of AGN Models on the \lya~P1D}
\shortauthors{Pirecki et al.}

\begin{document}

\title{Exploring the impact of AGN feedback model variations on the Lyman-$\alpha$~Forest Flux Power Spectrum}

\author[0009-0003-6251-3231]{Megan Pirecki}
\affiliation{Department of Physics and Astronomy, Rutgers University,  136 Frelinghuysen Rd, Piscataway, NJ 08854, USA}

\author[0000-0002-1185-4111]{Megan Taylor Tillman}
\affiliation{Department of Physics and Astronomy, Rutgers University,  136 Frelinghuysen Rd, Piscataway, NJ 08854, USA}

\author[0000-0001-5817-5944]{Blakesley Burkhart}
\affiliation{Department of Physics and Astronomy, Rutgers University,  136 Frelinghuysen Rd, Piscataway, NJ 08854, USA}
\affiliation{Center for Computational Astrophysics, Flatiron Institute, 162 Fifth Avenue, New York, NY 10010, USA}

 \author[0000-0002-8710-9206]{Stephanie Tonnesen}
 \affiliation{Center for Computational Astrophysics, Flatiron Institute, 162 Fifth Avenue, New York, NY 10010, USA}

 \author[0000-0001-5803-5490]{Simeon Bird}
 \affiliation{University of California, Riverside, 92507 CA, U.S.A.}

\begin{abstract}
We study the effects of varying different Active Galactic Nuclei (AGN) feedback parameters on the Lyman-$\alpha$~(Ly$\alpha$) forest 1D transmitted flux power spectrum (P1D). 
We use the Cosmological and Astrophysics with Machine Learning Simulations (CAMELS) suite to explore variations on the Simba simulation AGN feedback model. 
The parameters explored include AGN momentum flux, AGN jet speed, supermassive black hole (SMBH)  radiative efficiency, jet velocity threshold, and minimum SMBH mass needed to produce jet feedback. 
Although all parameters affect the P1D, this work explores the radiative efficiency, jet velocity threshold, and minimum SMBH mass in this context for the first time and finds the following results:
Primarily, the most massive SMBHs impact the Ly$\alpha$~forest through the jet feedback mode.
While heating AGN jets to the virial temperature at injection aids in the removal of neutral hydrogen from the Ly$\alpha$~forest, this heating also inhibits further jet feedback.
Similar behaviors are seen when varying the SMBH radiative efficiency, with higher values resulting in a suppression of SMBH growth and thus a later reduction in AGN feedback and lower values directly reducing the impact of AGN feedback on the Ly$\alpha$~forest P1D. 
These results imply that increasing the AGN feedback strength in the Simba simulation model suppresses the Ly$\alpha$~forest P1D, but only if the feedback does not impact the number of massive jet producing BHs. 
Future studies of AGN feedback models will require careful exploration of the unique aspects of the specific subgrid model, and how they interact with one another, for a complete understanding of the potential astrophysical impacts of SMBH feedback.

\end{abstract}


\section{Introduction}\label{s:Introduction}
The majority of baryonic matter in the Universe exists in the vast regions between galaxies and galaxy clusters, referred to as the intergalactic medium (IGM). The IGM can be broadly divided into two categories: cool, diffuse gas (with temperatures $\sim 10^4$ K) and warm-hot intergalactic medium \citep[WHIM;][]{Dave2001}. Current observations show that about $30\pm10$\% of the gas in the Universe is observable in the form of \lya~absorption (HI), which is predominantly found in the cooler, diffuse phase of the IGM \citep{McQuinn:2016}. 
At $z>1.8$, the \lya~forest is visible through ground-based telescopes, resulting in a rich collection of observational data, supplemented by ongoing surveys \citep[e.g.\ from BOSS, eBOSS, SDSS III and IV, Keck/HIRES, and VLT/UVES;][]{Vogt:1994HIRES, Dekker:2000UVES, Viel:2008, Eisenstein:2011SDSSIII, Dawson:2013BOSS, Viel:2013aKeck/HIRES, Dawson:2016eBOSS, Pieri:2016WEAVE-SQO, DESI:2016, Blanton:2017SDSSIV, Walther:2018VLT/UVES}.
An effective tool for cosmology and astrophysics, the \lya~forest can be used to investigate alternative dark matter models and determine cosmological parameters \citep{Viel+Haehnelt:2006,Armengaud:2017, Irsic:2017, Bird:2023, Fernandez:2023}.  The \lya~forest is also sensitive to the temperature evolution of the diffuse IGM from reionization to the present day, since the IGM temperature is expected to increase from $T_0\sim7500$ K at $z\sim 4.5$ \citep{Boera:2019} to $T_0\sim14750$ K due to HeII reionization at $z\sim 3$ \citep{Gaikwad:2021}, before dropping to around $T_0\sim 5000$ K by $z=0$ \citep{McQuinn:2016, UptonSanderbeck:2016}.

In recent years, there has been a growing interest in understanding how galactic feedback, particularly from active galactic nuclei (AGN), impacts the low-$z$ \lya~forest \citep{Viel:2017, Gurvich2017, Burkhart_2022, Tillman:2023AJL, Tillman:2023AJ, Dong:2023, Dong:2024, Tillman:2025}. Feedback models are often tuned to fit the observed galaxy stellar mass function  \citep{dave:2019, Pillepich:2018a, McCarthy_2017, Crain_2015, Schaye_2014}, and various AGN models have produced galaxies that align well with these observations. However, additional observational constraints will be necessary to differentiate between AGN feedback models, and the \lya~forest offers a promising avenue for this.

Numerous studies have attempted to explore how AGN feedback affects the IGM at low redshifts, but a clear method to disentangle the effects of AGN from other sources of heating and ionization remains elusive \citep{Nasir:2017, Christiansen:2020, Chabanier:2020, Burkhart_2022, Khaire:2022, Tillman:2023AJL, Tillman:2023AJ, Dong:2023, Khaire:2023, Dong:2024, Khaire:2024}. The effect of AGN feedback on the low-$z$ column density distribution function is largely degenerate with the effect of scaling the amplitude of the metagalactic ultraviolet background \citep[UVB,][]{Khaire:2015,Burkhart_2022,Tillman:2023AJL,Khaire:2024}. Intriguingly, no existing simulation reproduces the observed $b$-value distribution at $z<1$ \citep{Viel:2017}. Among the various \lya~forest summary statistics, \citet{Burkhart_2022} and \citet{Khaire:2024} identified the 1D transmitted flux power spectrum (P1D) as the most promising tool to uncover the unique aspect of AGN feedback.
Despite the fact that the effects of AGN feedback and the UVB on the P1D are degenerate to some extent, the P1D probes unique mechanisms at its different scales.
This is because while the UVB will impact the P1D on all scales, AGN heating, that impacts individual \lya~forest absorbers, will most strongly affect small physical scales.

On small scales and at high redshift ($z>2$), the P1D provides valuable information on the thermal state of the IGM, which is essential to understand reionization \citep{Zaldarriaga:2001, Meiksin:2009, Lee:2015, McQuinn:2016, Boera:2019, Zhu:2021} and the thermal history of the IGM \citep{Viel+Haehnelt:2006, Bolton:2008, Becker:2011, Rudie:2012, Hu:2023}. 
Moreover, the P1D, by tracing small-scale structures ($k \sim 0.2 - 20$ h/Mpc), can constrain cosmological parameters such as neutrino mass \citep{Palanque-Delabrouille:2015a, Palanque-Delabrouille:2015b, Yeche_2017, Palanque-Delabrouille:2020}, the nature of dark matter \citep{Yeche_2017,Palanque-Delabrouille:2020,Villasenor:2023, Baur:2016}, and extensions to the cosmological standard model such as dark radiation and sterile neutrinos \citep[e.g.][]{Baur:2017, Rossi:2017, Garny:2018, Rossi:2020}. 

In particular, some works have found that AGN feedback could have an impact on the predicted P1D at $z>2$ \citep{Viel:2009,Chabanier:2020}, although these models were not calibrated to match the observed star formation rate.
More recent works on state-of-the-art AGN feedback models found that AGN feedback effects were smaller than previous works predicted at high redshift \citep[$z>2$][]{Bird:2023, Tillman:2025}, but began to affect the P1D at the 10\% level at lower redshifts ($z<1$). 
To understand how AGN feedback affects the \lya~forest, exploring the impact of more detailed variations of AGN feedback models on the P1D is crucial.

In this paper, we explore the effects of AGN feedback on the \lya~forest P1D.
We do so by analyzing the variations in the Simba AGN feedback model using CAMELS simulations \citep{SIMBA, CAMELS-public}. 
From the CAMELS simulations we generate synthetic spectra to analyze the P1D across a wide range of redshifts ($z=0.1-2.0$).
The paper is outlined as follows. In Section~\ref{s:Methods} we describe the simulations analyzed and how the synthetic spectra are generated and processed.
In Section~\ref{s:SimbaP1D} we present the predicted P1D from the parameter variations of interest in several redshift bins.
In Section~\ref{s:Discussion} we discuss the results, why the parameters have an impact on the P1D, and how these results fit into the broader discussion of AGN feedback impacts on IGM scales.
Finally, in Section~\ref{s:Conclusion} we present our conclusions and discuss the next steps motivated by this work.

\section{Methodology}\label{s:Methods}
    
    \subsection{Simulations}\label{ss:simulations}
        This work uses data from the Cosmology and Astrophysics with MachinE Learning Simulations (CAMELS) project \citep{camels, Ni:2023}. 
        CAMELS consists of thousands of N-body and (magneto)hydrodynamical simulations run with the AREPO, GIZMO, and GADGET codes \citep{Springel:2005,Springel:2010,Hopkins:2015, Hopkins:2017,Weinberger:2020}. 
        The CAMELS project includes simulations based on IllustrisTNG \citep{IllustrisTNG}, Simba \citep{SIMBA}, and Astrid \citep{Bird:2022}, with further models under development.
        For each suite of simulations in CAMELS, there is a set of simulations that vary a single parameter out of a total of 28.
        These parameters correspond to different aspects of the cosmological and astrophysical subgrid models. 
        Analyzing these single-parameter variations allows for the isolation and study of their overall impact.
        Additionally, each CAMELS suite uses the same baseline cosmology values to further simplify comparisons between models.
        For each simulation used in this paper the cosmological parameters are $\Omega_b = 0.049$, $h = 0.6711$, $n_s = 0.9624$, $M_\nu = 0$ eV, $w=-1$, $\Omega_k=0$.

        As this work aims to further explore the AGN feedback model in the Simba simulations \citep[following work done in][]{Tillman:2023AJ}, we analyze only the CAMELS-Simba simulations that vary different aspects of the AGN feedback model.
        There are five parameters that are directly related to the AGN feedback model.
        For each of these parameters, there are CAMELS simulations that increase and decrease the value to study the effect. 
        These parameters are as follows, as also seen in Table~\ref{tab:summary}: 
        
        (1) $A_\textnormal{AGN1}$ scales the overall AGN feedback momentum flux. 
        Both the radiative and jet AGN feedback modes are implemented kinetically and $A_\textnormal{AGN1}$ varies the total momentum flux of these outflows.
        This value is sampled between 0.25 and 4 with the fiducial value being $A_\textnormal{AGN1}$ = 1.

        (2) $A_\textnormal{AGN2}$ scales the overall speed of the AGN jets.
        In the jet mode, outflows have an additional velocity component, dependent on the Eddington ratio, that can reach up to 7000 km/s in the fiducial case. 
        $A_\textnormal{AGN2}$ scales this maximum value by a factor of 0.5 to 2 with the fiducial value of $A_\textnormal{AGN2}$ = 1.
        
        (3) BHRadiativeEff corresponds to the radiative efficiency of the BH accretion disk. Varying this value changes the fraction of accreted material that is returned as energy injected through feedback. BHRadiativeEff varies from 0.025 to 0.4 with a fiducial value of 0.1.
        
        (4) BHJetTvirVel sets the velocity that an AGN jet must have to be heated to the virial temperature of the host. 
        In the fiducial model, AGN jets with a velocity of at least 2000 km/s are heated to $T_{vir}$ at ejection. BHJetTvirVel varies from 500 km/s to 8000 km/s.
        
        (5) BHJetMassThr changes the mass threshold above which SMBHs can produce jet feedback. This ranges from $4.5 \times 10^6M_\odot$ to $4.5 \times 10^8M_\odot$ with fiducial value $4.5\times10^7M_\odot$.
       
        For each parameter, we will compare the predictions of the \lya~forest P1D from the highest, lowest, and fiducial value.
        This provides a clearer picture on the impact that the individual feedback components in the model have. 
        For more information on the Simba simulation subgrid models, see \citet{dave:2019}.
        The implementation of parameter variations is described in the CAMELS public data release papers \citep{CAMELS-public, Ni:2023}. 
        Furthermore, we summarize the parameters explored herein and the range of values explored in Table~\ref{tab:summary}.

    \nolinenumbers
    \begin{table*} \caption{Table summarizing AGN feedback parameters explored in CAMELS-Simba.}\label{tab:summary}
        \begin{tabularx}{\textwidth}{lP{0.4\textwidth}ccc} 
            \hline\textbf{Parameter} & \textbf{Description} & \textbf{Low Value} & \textbf{Fiducial} & \textbf{High Value} \\ \hline
            $A_\textnormal{AGN1}$ & Normalization factor for the momentum flux of AGN. & 0.25 & 1.0 & 4.0 \\ \hline
            $A_\textnormal{AGN2}$ & Normalization factor for the maximum speed of AGN jets. & 0.5 & 1.0 & 2.0 \\ \hline
            BHRadiativeEff & Radiative efficiency of the black hole accretion disk. & 0.025 & 0.1 & 0.4 \\ \hline
            BHJetTvirVel & Outflow velocity above which gas ejected is heated to the virial temperature. & 500 km/s & 2000 km/s & 8000 km/s \\ \hline
            BHJetMassThr & Black hole mass above which the jet mode can be active. &$4.5\times10^6\ M_\odot$  & $4.5\times10^7 M_\odot$  & $4.5\times10^8 M_\odot$ \\ \hline
        \end{tabularx}
        \end{table*}

    \subsection{Synthetic Spectra}\label{ss:Spectra}
    
        Synthetic spectra are generated from the CAMELS simulations using Fake Spectra Flux Extractor (FSFE)\footnote{\url{https://github.com/sbird/fake_spectra}}, a well-tested public python / C++-based code described in \citet{Bird:2015, Bird:2017} with MPI support from \citet{Quezlou:2022}. 
        The code places random sightlines in the simulation box and generates spectra for the ion of choice.
        To minimize the effect of sample variance on the P1D, 5,000 random sightlines are generated.
        We refrain from adding noise to the spectra to simulate a high signal-to-noise ratio limit.
        Spectral lines are generated by interpolating the neutral hydrogen mass in each gas element onto the sightline using a smoothed particle hydrodynamics kernel. Specifically, the CAMELS-Simba simulations use a cubic spline kernel.
        Spectra are generated with a pixel width of 1 km/s.
        We define flux $\mathcal{F}$ as a function of optical depth $\tau$ using the relation $\mathcal{F}(v) = \exp{(-\tau)}$.

    \subsection[The 1D Lyman alpha Flux Power Spectrum]{The 1D \lya~Flux Power Spectrum}\label{ss:FPS}
    
        From our generated spectra we calculate the predicted 1D \lya~forest flux power spectrum (P1D).
        We find the 1D Fourier transform of the flux fluctuations $\delta_{\mathcal{F}}\equiv \mathcal{F}(v)/ \langle \mathcal{F}\rangle - 1$, and average them over all sightlines to find the P1D along the line of sight.
        The mean transmitted flux $\langle \mathcal{F} \rangle$ is computed over all sightlines in a given redshift bin. 
        The resulting power spectrum ranges from wavenumbers of $k \sim 3\times10^{-3}$ s/km (corresponding to the simulation co-moving box length of 25 Mpc/h) to $k \sim 0.1$ s/km, the smallest $k$ probed by current observations.
        
        We acknowledge that on the smallest scales, the P1D predicted by the CAMELS simulations is underresolved, as was the case for the column density distribution function and the $b$-value distribution \citep{Tillman:2023AJ}.
        The impacts of a lack of convergence on the \lya~forest P1D were discussed in detail in the Appendix of \citet{Tillman:2025}.
        The relatively small size of the CAMELS simulation boxes means that the P1D will be significantly affected by cosmic variance, as the halo mass function will not be fully sampled in a 25 Mpc/h box.
        Since the impact of AGN feedback will depend on the specific halo sample, we expect the quantitative results of this study to vary in a larger simulation box.
        Therefore, we refrain from comparing directly to observations due to the expected lack of convergence and instead focus on the trends when varying the parameter values.
        This should pose no issues, as the box size, resolution, and initial random seed are maintained between all the simulations compared within.

\section{Simba P1D}\label{s:SimbaP1D}

We will now present the impact of varying the CAMELS-Simba AGN parameters on the \lya P1D. Each parameter variation is represented in its own figure that presents the P1D for the minimum, fiducial, and maximum value of the shown parameter. 
For each figure, the top row displays the P1D while the bottom row displays the ratio of the P1D of parameter variation simulations to the P1D of the fiducial simulation. 
The red lines represent the maximum value for the parameter variation, the blue lines show the minimum value, and the black lines show the fiducial results. 
In the ratio plots, a gray dashed line is included to represent a difference of 10\% from the fiducial simulation.
The ratios provide a clearer picture of the changes in the P1D caused by the parameter variations. 
Each figure includes redshift values from 0.1 to 2.0. 
Specifically, we used $z=0.1$, $0.4$, $0.7$, $1.0$, $1.25$, and $2.0$. 

The CAMELS-Simba simulations vary a total of 28 parameters \citep{Ni:2023}; however, only 5 of the parameters are analyzed herein. 
We explore CAMELS-Simba parameter variations that change aspects of AGN feedback specifically such as jet threshold mass, jet speed, and momentum flux. 

\begin{figure*}
                \centering
                \includegraphics[width = \linewidth, trim=0.0cm 0.0cm 0.0cm 0.0cm, clip=true]{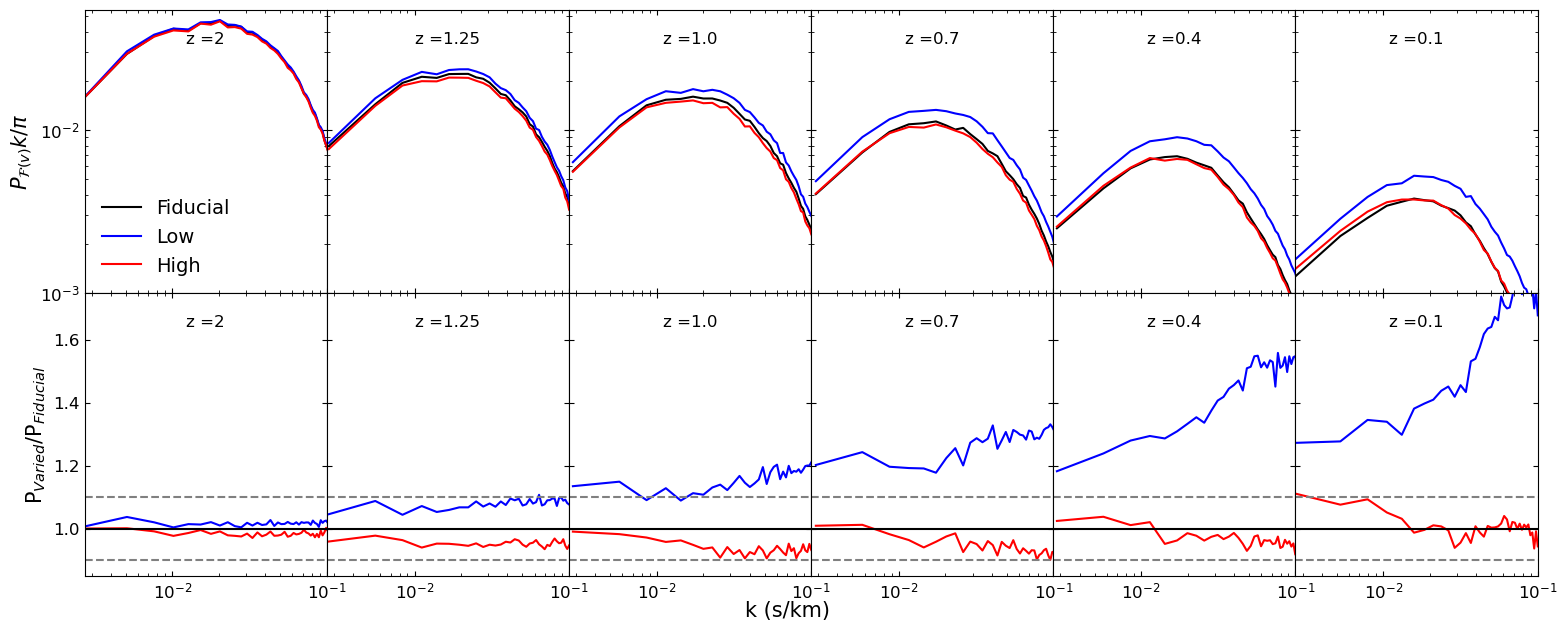}
                \caption{The \lya~forest transmitted 1D flux power spectrum for different redshifts varying the SMBH momentum flux ($A_{AGN1}$). The top row is the P1D and the bottom row shows the ratios of the P1D relative to the fiducial. The black line is the fiducial simulation value while red corresponds to a high parameter value, and blue is the low parameter value. The dashed grey line is the 10\%\ difference from the fiducial simulation. Only decreasing the SMBH momentum flux from the fiducial value results in a significant change in the PID (difference of more than 10\%)}
                \label{fig:flux}
            \end{figure*}

Figure \ref{fig:flux} shows the P1D for variations in the AGN momentum flux. Increasing the momentum flux does not change the P1D more than 5\% at any $z$. 
Decreasing the momentum flux causes small (less than 10\%) changes in the P1D for $z > 1$, but at $z=1$ the P1D is enhanced by more than 50\% and at $z=0.1$ enhanced up to 80\%. 
The largest change in the P1D occurs for $k \gtrsim 3\times 10^{-2}$ s/km.
As the redshift decreases, the relative change in the P1D for low vs.\ high $k$ values becomes more dramatic.

\begin{figure*}
                \centering
                \includegraphics[width = \linewidth, trim=0.0cm 0.0cm 0.0cm 0.0cm, clip=true]{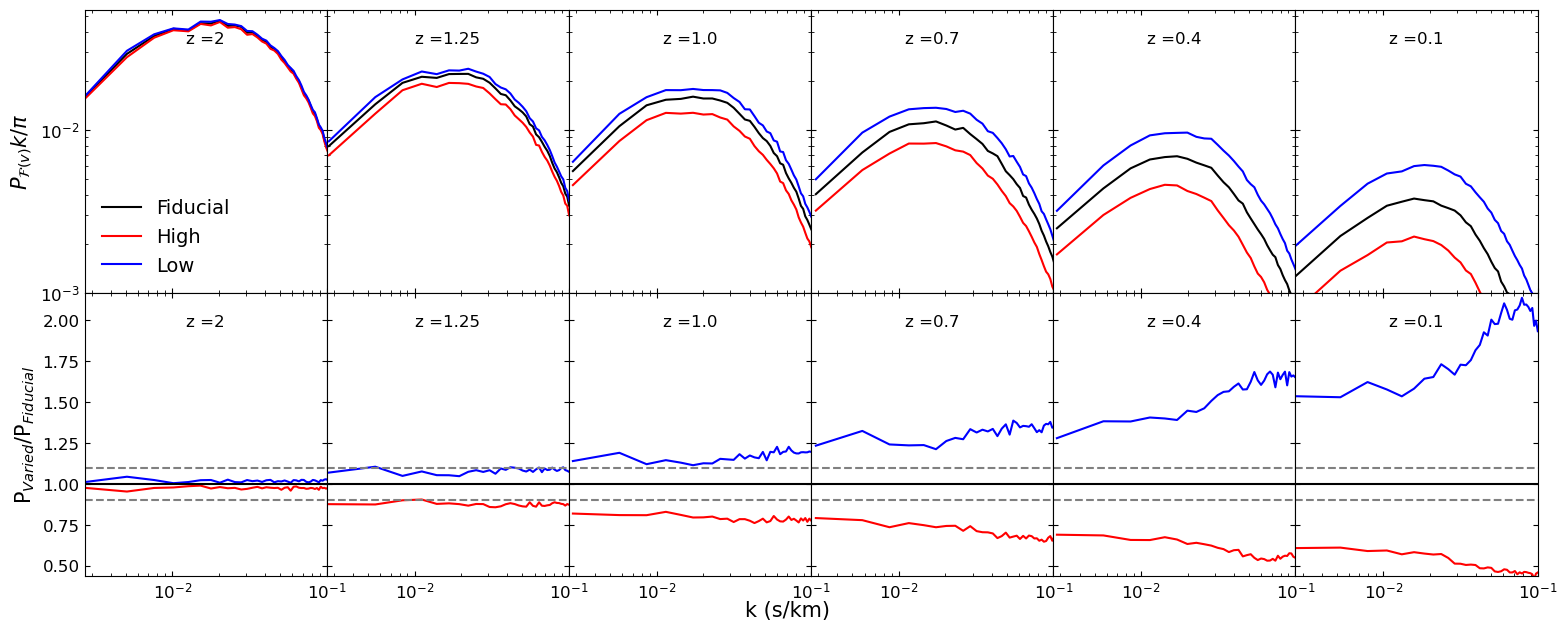}
                \caption{Same layout as Figure 1 but now showing variations in the maximum AGN jet speed parameter ($A_{AGN2}$). The top row shows the P1D and the bottom row shows the P1D ratios to the fiducial simulation. The colored lines are variations from the fiducial value. Both increases and decreases to AGN jet speed cause dramatic changes at low $z$.}
                \label{fig:jet_speed}
            \end{figure*}

Figure \ref{fig:jet_speed} shows the impact of varying the maximum speed of the AGN jets on the predicted P1D. 
Increasing the jet speed depresses the P1D, and decreasing the jet speed enhances the P1D.  
Around $z=1.0$, when the value of $A_{AGN2}$ varies in either direction, the changes in the P1D become greater than 10\% and by $z=0.1$ these changes are greater than 50\%. 
Decreasing the jet speed results in an increase in power up to 50-100\% by $z=0.1$. 
Increases in jet speed result in a decrease in power up to 50\%, and changing the jet speed has a stronger relative impact on the P1D at higher $k$. 

\begin{figure*}
                \centering
                \includegraphics[width = \linewidth, trim=0.0cm 0.0cm 0.0cm 0.0cm, clip=true]{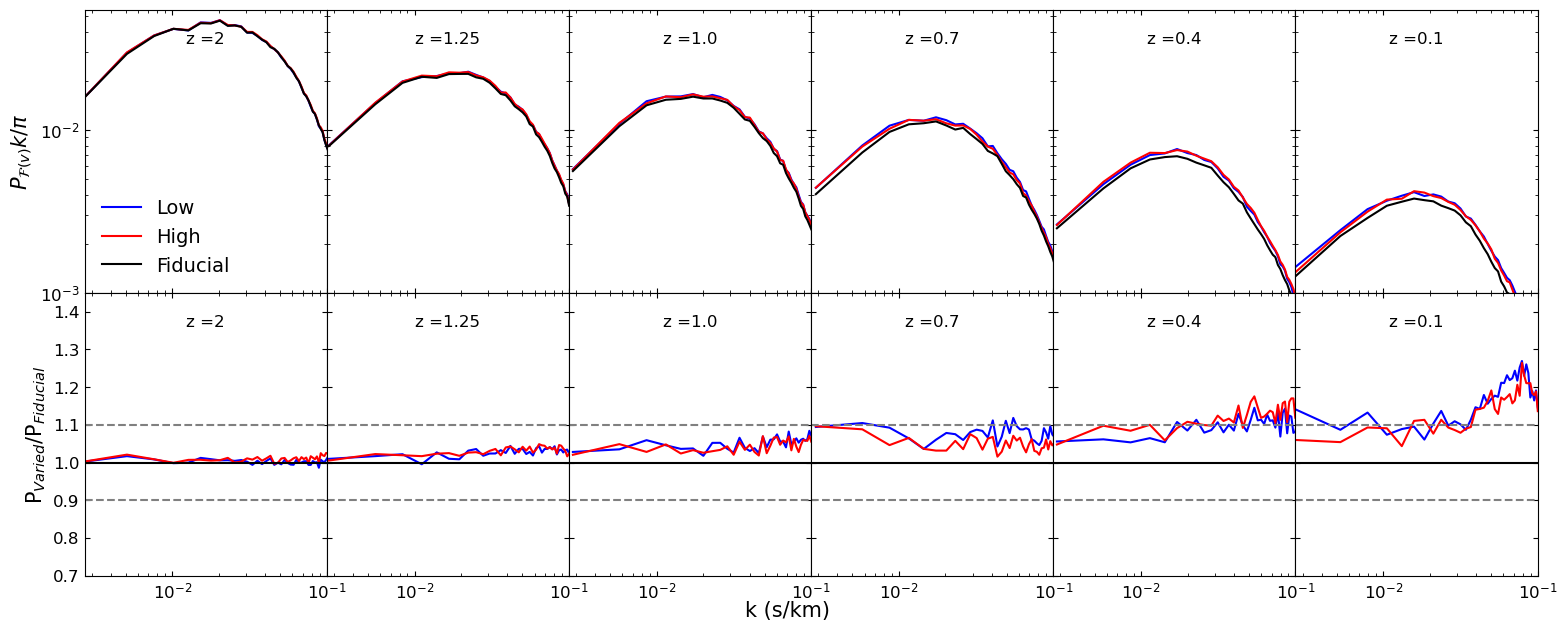 }
                \caption{Same layout as Figure 1 but now showing variations in the radiative efficiency of the black hole (BHRadiativeEff) parameter. The top row shows the P1D and the bottom row shows the P1D ratios to the fiducial simulation. The colored lines are variations from the fiducial value. Both increases and decreases of BHRadiativeEff results in a similar gain of power.}
                \label{fig:1P_26}
            \end{figure*}

\begin{figure*}
                \centering
                \includegraphics[width = \linewidth, trim=0.0cm 0.0cm 0.0cm 0.0cm, clip=true]{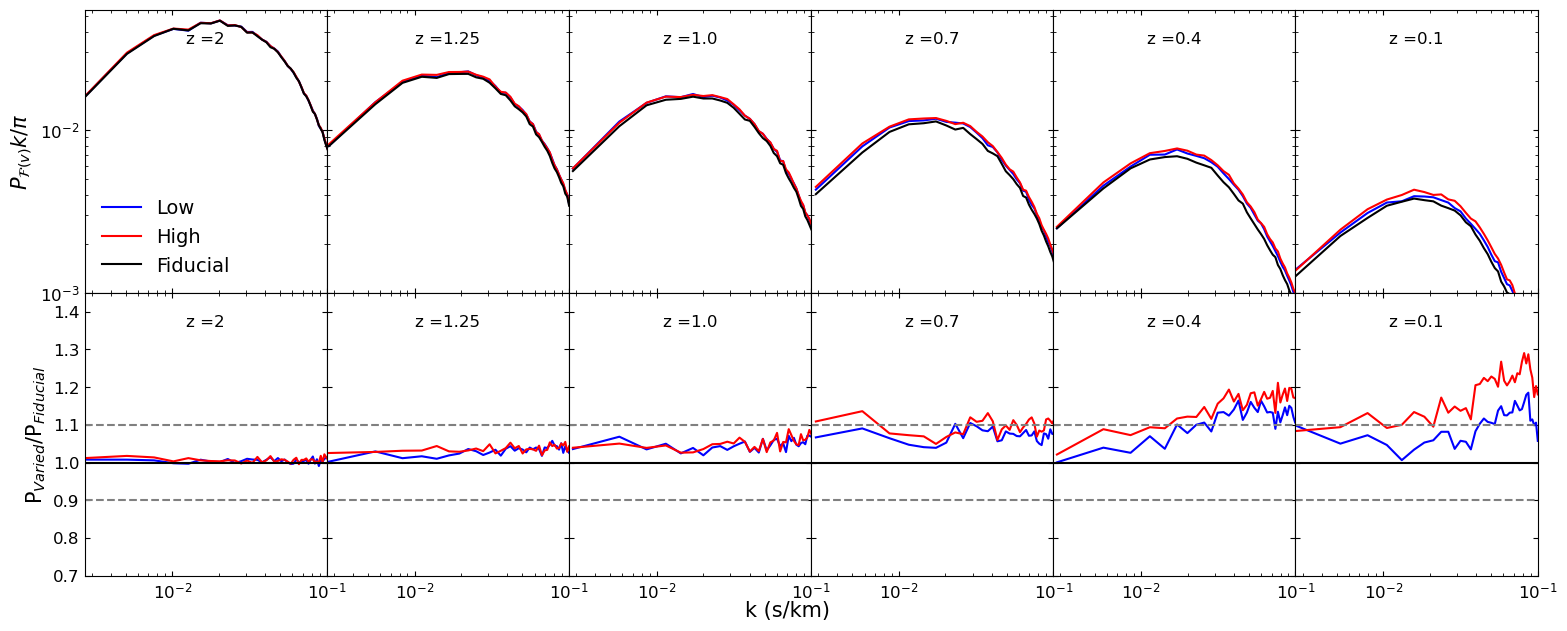 }
                \caption{Same layout as Figure 1 but now showing variations in the SMBH jet velocity threshold at which the jet is heated to the virial temperature (BHJetTvirVel). The top row shows the P1D and the bottom row shows the P1D ratios to the fiducial simulation. The colored lines are variations from the fiducial value. Any changes in BHJetTvirVel result in a gain of power, but increasing BHJetTvirVel proves to be more impactful than decreasing.}
                \label{fig:1P_27}
            \end{figure*}

Figures \ref{fig:1P_26} and \ref{fig:1P_27} show variations in the SMBH radiative efficiency (BHRadiativeEff) and the jet velocity threshold for heating (BHJetTvirVel), respectively.
We first note that varying these parameters appears to have a smaller effect on the P1D than $A_{AGN1}$ and $A_{AGN2}$, only crossing the 10\% threshold at $z\leq 0.4$.  Interestingly, for both BHRadiativeEff and BHJetTvirVel, increasing or decreasing the fiducial value leads to an enhanced P1D. We will discuss this in more detail in Section \ref{s:D:ss:other}.

\begin{figure*}
                \centering
                \includegraphics[width = \linewidth, trim=0.0cm 0.0cm 0.0cm 0.0cm, clip=true]{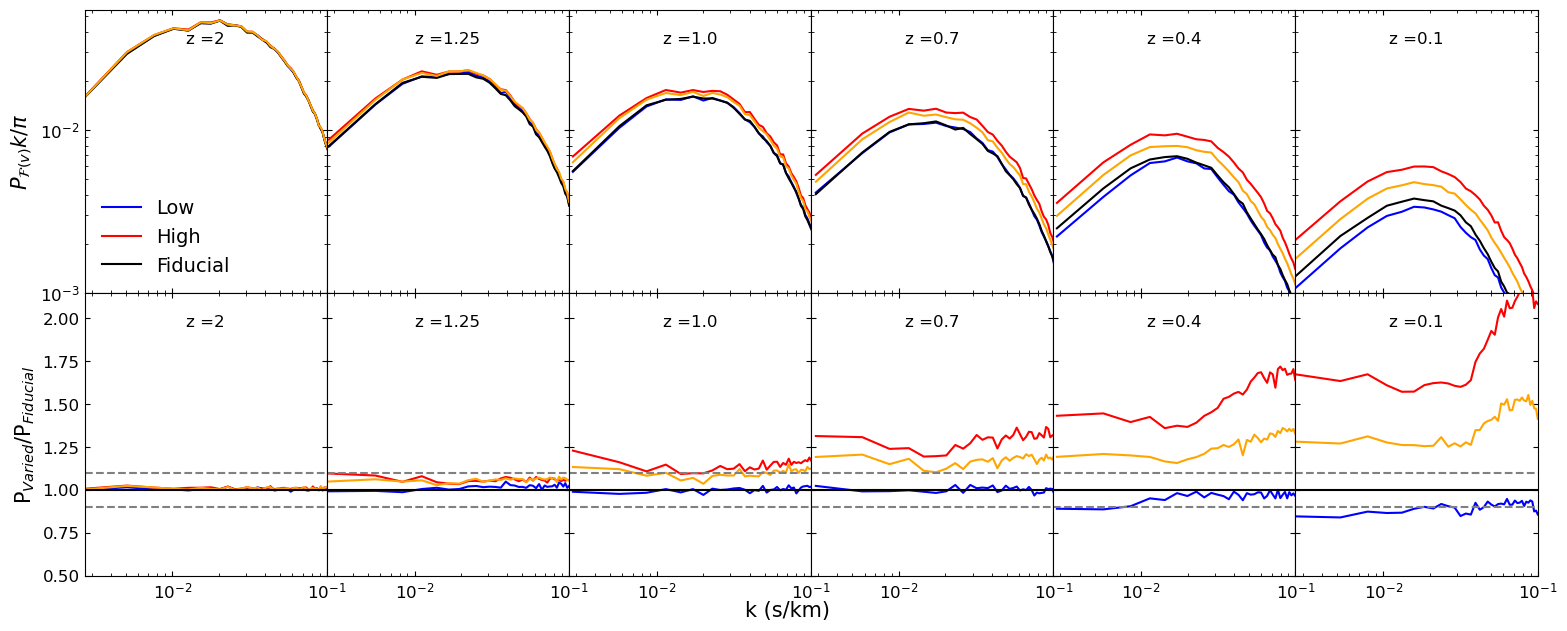 }
                \caption{Same layout as Figure 1 but now showing variations in the minimum SMBH mass required to produce jet feedback (BHJetMassThr) parameter. The top row shows the P1D and the bottom row shows the P1D ratios to the fiducial simulation. The colored lines are variations from the fiducial value. The orange line corresponds to the predicted P1D with a BHJetMassThr value between fiducial and high (BHJetMassThr = 4.5$\times 10^{7.5} M_\odot$). Only increases in BHJetMassThr result in a substantial impact implying that the most massive SMBHs are those affecting the forest.}
                \label{fig:1P_28}
            \end{figure*}

Figure \ref{fig:1P_28} shows the variation in the predicted P1D when changing the threshold mass (BHJetMassThr) of the AGN jet mode. 
Increases in the jet threshold mass cause an increase in power, while decreases in the mass cause only a small decrease in power.
For $z > 1.0$, changes in the P1D when the jet threshold mass is increased are less than 10\%.  
Around $z=0.1$, the P1D shows changes of 50-100\%, and the change is greater with higher k.
Decreasing the threshold mass causes a 10\% loss of power by $z=0.1$ but changes to the P1D at $z>0.4$ are negligible. The orange line represents the predicted P1D with a BHJetMassThr value halfway between fiducial and high ($4.5\times 10^{7.5} M_\odot$). 
The resulting change to the P1D at this parameter value lies about halfway between the predictions for fiducial and high values of the parameter.

            \begin{figure*}
                \centering
                \includegraphics[width = 0.75\linewidth, trim=0.0cm 0.0cm 0.0cm 0.0cm, clip=true]{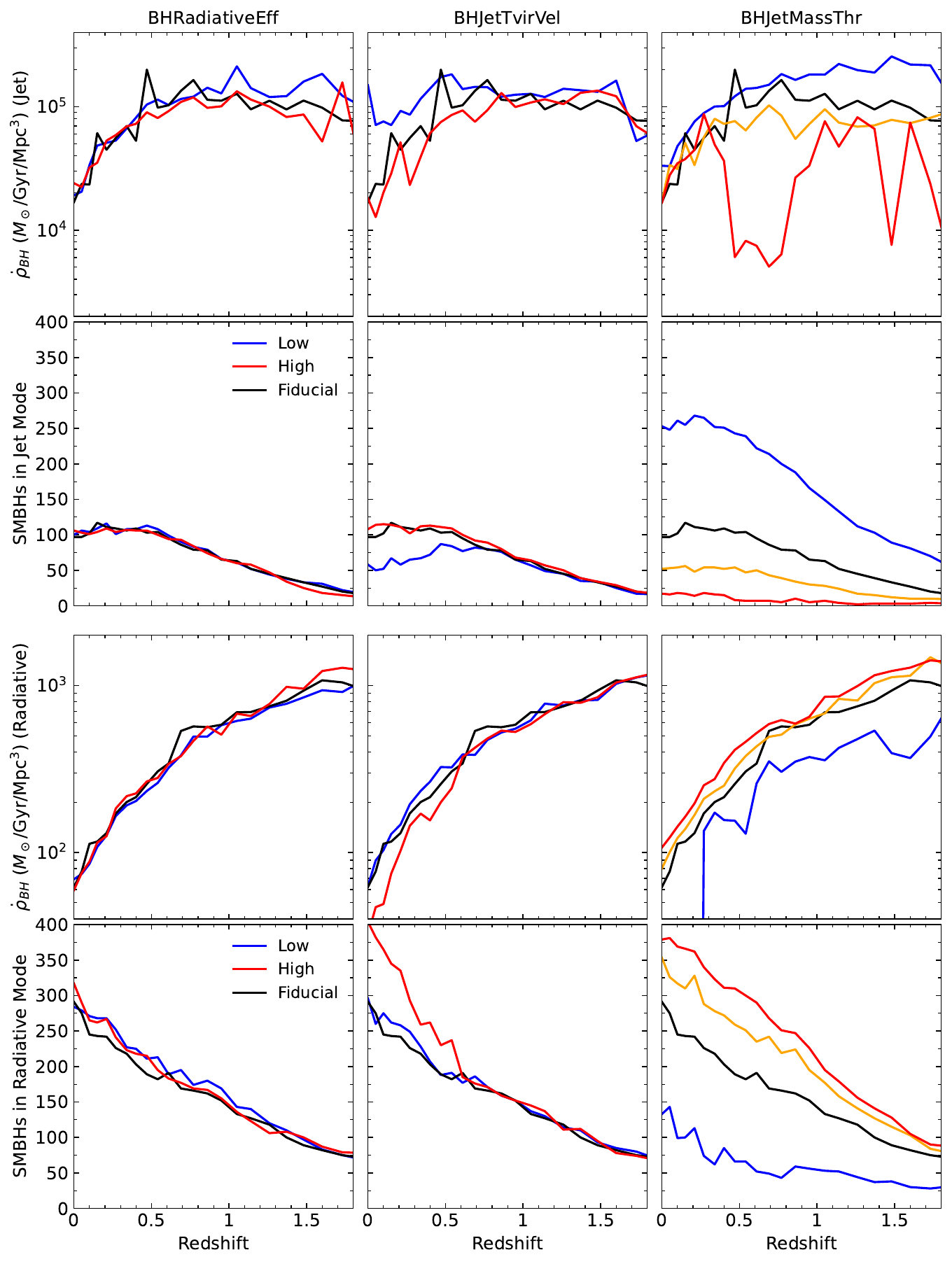 }
                \caption{The accretion rate density of SMBHs and the total number of SMBHs for both the radiative mode (top two plots) and the jet mode (bottom two plots) as a function of redshift for the BHRadiativeEff, BHJetTvirVel, and BHJetMassThr parameters. Similar data for $A_{AGN1}$ and $A_{AGN2}$ is presented in \citet{Tillman:2023AJ}. The colored lines correspond to the low, fiducial, and high values for each parameter. For BHJetMassThr, there is an additional orange line which denotes a value between fiducial and high. The top right panel shows a sharp drop for high BHJetMassThr. This is a result of the low fraction of SMBHs in the jet mode; the accretion rate density is \textbf{noisy} since only a few SMBHs contribute in this case compared to the low BHJetMassThr and fiducial cases.}
                \label{fig:BHstats}
            \end{figure*}

\section{Discussion}\label{s:Discussion}

We have presented how changing the AGN feedback parameters within the CAMELS-Simba simulations impacts the predicted P1D of the \lya~forest. In this section, we will discuss the broader implications of how the variation of these parameters physically affects the \lya~forest.  
To put our results in a larger context, we summarize, from previous work, how AGN feedback is expected to impact the \lya~forest. 
\citet{Tillman:2025} found that jet feedback reduces power in the \lya~forest and argued that this is because the jets heat the surrounding medium.  
Therefore, we posit that any feedback parameter that results in a higher fraction of galaxies with AGN jets will heat more of the IGM and cause a loss in power as seen through the P1D.  
In addition, jets that travel further into the IGM should heat more of the IGM and reduce the P1D \citep[as seen in][when varying AGN jet speed]{Tillman:2023AJ}.
Finally, we might expect that the introduction of more hot material into the IGM from the AGN feedback will result in the overall heating of the IGM through mixing. 
Changes at the high $k$ end of the P1D ($k \gtrsim 0.03$ s/km) can be the result of heating through the broadening of individual absorbers in the vicinity of a massive galaxy hosting an AGN.
Changes at the low-$k$ end primarily reflect changes in the total ionization fraction of hydrogen in the \lya~forest.
With these mechanisms in mind, we will discuss the parameters individually below.

The AGN jet mode has been shown to have the greatest impact on the \lya~forest \citep{Tillman:2023AJ}, although the radiative mode can affect the \lya~forest as well.
In general, when more SMBHs produce jet feedback, there is a greater loss of power from the \lya~forest.
To examine the relative impacts of the different AGN feedback modes, we analyze SMBH statistics in each feedback mode individually.
Figure~\ref{fig:BHstats} shows the accretion rate density of SMBHs in the different feedback modes, as well as how the total number of SMBHs in each mode across redshift depends on the BHRadiativeEff, BHJetTvirVel, and BHJetMassThr parameters. 
Similar results are provided for \AAGN{1} and \AAGN{2} in \citet{Tillman:2023AJ} and these results help explain some of the impacts of different parameters on the P1D.
Note that the accretion rate densities for the jet mode are a factor of 100 greater than that of the radiative mode.
Therefore, in general, we should expect changes in the jet mode to have a greater impact on the simulation box than changes to the radiative mode.

\subsection{Impact of the SMBH momentum flux and the maximum AGN jet velocity on the neutral IGM}
    The impact of variations in momentum flux and maximum AGN jet velocity (\AAGN{1} and \AAGN{2}) on the \lya~forest column density distribution and $b$-value distribution have already been analyzed in \citet{Tillman:2023AJ}. 
    Impacts on the P1D as seen herein resemble the results seen in previous work; namely the important impact of AGN jet mode feedback on the properties of the \lya~forest. 
    The results when varying \AAGN{2} as seen in Figure \ref{fig:jet_speed}, the normalization factor for the speed of the AGN jets, are the most intuitive, as the AGN jets are the main culprit in \lya~forest heating.
    A decrease in \AAGN{2} means that the jets cannot propagate through the simulation box as far, resulting in less heating and thus more neutral hydrogen.
    Increasing \AAGN{2} allows the jets to travel farther, heating more of the IGM, which causes a decrease in neutral hydrogen.
    Decreasing \AAGN{2} will have the opposite effect.
    However, the impact of increasing \AAGN{2} is not as drastic as that of decreasing \AAGN{2}.
    Furthermore, for increases and decreases in \AAGN{2}, the impact at high $k$ (corresponding physically to small scales) is greater since the jets are seen to impact both the thermal history of the IGM and the thermal broadening of \lya~absorbers \citep{Tillman:2025}.

    The results for the variations in \AAGN{1} are more complex.
    The decrease in \AAGN{1} shows clear increases in power, while the increase in \AAGN{1} does not affect the P1D more than 10\%. 
    This implies that the magnitude of the effects of increasing the AGN momentum flux have peaked at the fiducial value.
    The momentum flux from the AGN feedback can effectively remove power from the \lya~forest and, since there are larger impacts at the high $k$ end of the P1D, this corresponds to a heating effect which can change the overall temperature of the cold-diffuse IGM.
    The impact at high $k$ also implies that the presence of this type of AGN feedback could thermally broaden individual \lya~absorbing structures.
    Results from \citet{Tillman:2023AJ} indeed show an impact on the $b$-value distribution, implying that the change at the high-$k$ ($k > 3\times10^{-2}$) end of the P1D is likely due to the heating of individual \lya~absorbers and the overall diffuse IGM.
    
\subsection{Impact of other SMBH feedback model parameters on the neutral IGM}\label{s:D:ss:other}
    
    The results from varying the SMBH radiative efficiency (BHRadiativeEff) and the jet velocity threshold at which AGN jets are heated (BHJetTvirVel) are similar. 
    Furthermore, these two parameters are the least impactful of the parameters explored in this work. 
    
    Varying BHRadiativeEff changes the amount of energy injected via the X-ray feedback mode and through the AGN momentum flux.
    Similarly to \AAGN{1}, this parameter is a normalization factor for the energy released through the radiative AGN feedback mode, but unlike \AAGN{1}, this parameter is also a normalization factor for the X-ray feedback mode.
    As such, decreases and increases in this parameter likely affects the P1D power similarly to the results seen when changing the AGN momentum flux (\AAGN{1}). 
    However, since the values for BHRadiativeEff are a factor of 10 smaller than that of \AAGN{1}, the effect would be diminished.
    Deviations from the expected behavior are likely due to the impact of the X-ray feedback mode.

\begin{figure*}[!ht]
    \centering
    \includegraphics[width = \linewidth, trim=0cm 0.0cm 0.0cm 1cm, clip=true]{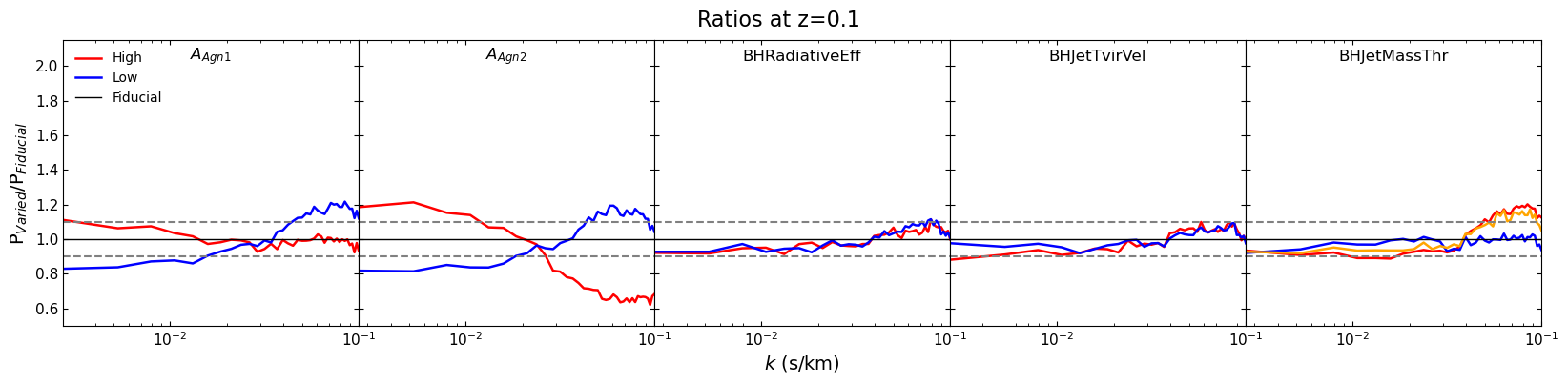 }
    \caption{Ratios of the \lya~forest P1D, with the mean transmitted flux corrected to match that of the fiducial simulation at $z=0.1$. For BHJetMassThr, there is an orange line which denotes a value between fiducial and high.}
    \label{fig:fig_combined}
\end{figure*}

    The decreases in BHRadiativeEff behave as expected given the impacts seen for \AAGN{1}. 
    However, increases in BHRadiativeEff cause a notable increase in power at high-$k$ not seen for \AAGN{1}.
    In \citet{Tillman:2025}, the addition of the X-ray feedback mode showed an increase in power at high-$k$ that was due to the impact of AGN feedback on the thermal evolution of the IGM.
    The X-ray feedback mode is unlikely to directly affect the IGM in such a way on its own because of its relatively local operation. 
    Therefore, the impact seen is probably due to the self-suppressive effect that the X-ray feedback mode has on the growth of the SMBHs, which would then impact the AGN feedback.
    In Figure~\ref{fig:BHstats}, minimal changes are seen in the accretion rate density and the number of SMBHs in the jet mode (upper two panels); therefore, the changes in the P1D must be mainly due to the changes seen in the number of SMBHs in the radiative feedback mode.
    This indicates that BHRadiativeEff does not affect the P1D via the most massive BHs in the simulations.
    However, a definitive claim can not be made due to limited statistics in our small-box simulations.

    BHJetTvirVel sets the velocity threshold at which the AGN jets ejected as feedback are heated to the virial temperature ($T_{vir}$) of the host halo.
    Jets with velocities above BHJetTvirVel will be heated to $T_{vir}$, while jets with lower velocities will have the same temperature as the gas originally accreted by the SMBH.
    Decreasing BHJetTvirVel allows slower jets to be heated.
    
    At $z=2$, the lowest value for BHJetTvirVel allows $\sim 20$\% more jets to be heated. Furthermore, minimal impact is seen on the SMBH accretion rate density and the number of AGNs that produce jet feedback in Figure~\ref{fig:BHstats}. At $z=2$, the overall number of SMBHs that produce AGN jet feedback is low enough so that additional jet heating does not affect the P1D in any significant way. Overall, the effect on the P1D is minimal at this redshift.
    By $z=0$, almost every jet is heated to $T_{vir}$ for both the minimal and fiducial value of BHJetTvirVel, therefore one might again expect little difference in the P1D.
    However, there are $\sim 30-40$\% fewer AGNs producing jet feedback in the low BHJetTvirVel case than there are for the fiducial case (see Figure~\ref{fig:BHstats}).
    The smaller number of jets at lower redshifts proves to be more important and results in more neutral hydrogen in the \lya~forest.
    
    When BHJetTvirVel is increased to its maximum value, the fraction of AGN jets heated to $T_{vir}$ drops to zero, resulting in a $\sim 10-30$\% increase in power for the P1D at $z=0.1$ (i.e.\ more neutral hydrogen in the \lya~forest).
    \cite{Tillman:2025} found an increase in power of $\sim 40-60$\% when completely removing jet feedback. 
    This implies that the impact of AGN jets on the \lya~forest is not solely dependent on the AGN jets being heated to the virial temperature, and, in fact, the majority of the impact is due to the kinetic aspect of the AGN jets alone.
    Thus, purely kinetic AGN jets would still have a dramatic impact on the predicted \lya~forest as the gas ejected is already at a higher temperature than the gas outside the galaxy, and the inclusion of additional heating only amplifies these effects.
    
    The final explored parameter was the minimum SMBH mass required to produce jet feedback (BHJetMassThr). 
    The larger the mass of the SMBH the more energy it is likely to release through jet feedback; however, so few SMBHs can reach the required mass that the impact of jet feedback is essentially reduced. 
    The larger impact at higher $k$ follows the impact that jet feedback has on both the thermal state of the IGM and the broadening of individual \lya~absorbers.
    Decreasing the mass required for an SMBH to produce jet feedback produces minimal change to the P1D.
    Smaller SMBHs will produce less energetic jets with lower velocities, meaning that their impact on the IGM will be lower.
    As seen in Figure~\ref{fig:BHstats}, the overall energy released in the jet mode is higher than that in the fiducial case, but not as high as expected, given the dramatic increase in jet-producing SMBHs.
    The smaller mass BHs must be accreting at lower rates, which will result in less momentum flux via AGN feedback.
    The impact of lowering BHJetMassThr by a factor of 10 from the fiducial value is less than half that of the impact of increasing BHJetMassThr by a factor of 5.
    These results imply that jets from SMBHs with masses of $M_{BH} > 10^{7} M_\odot$ are the main contributors to the impact we see on the \lya~forest from jet feedback.

\begin{table}[]\caption{Table displaying the maximum and minimum amplitudes of the \lya~P1D ratios show in Figure~\ref{fig:fig_combined} and the $k$ value at which the ratios cross 1.}
\label{tab:P1Dratiovalues}

\centering
\begin{tabular}{llll}
\hline
\textbf{Parameter}  & \textbf{$A_\textnormal{max}$} & \textbf{$A_\textnormal{min}$} & \textbf{$k_0$} \\ \hline
$A_\textnormal{AGN1}$ - High             & 1.07         & 0.92         & $1.4\times10^{-2}$       \\ \hline
$A_\textnormal{AGN1}$ - Low              & 1.22         & 0.84         & $3.5\times10^{-2}$       \\ \hline
$A_\textnormal{AGN2}$ - High             & 1.21         & 0.62         & $2.0\times10^{-2}$       \\ \hline
$A_\textnormal{AGN2}$ - Low              & 1.19         & 0.81         & $3.6\times10^{-2}$       \\ \hline
BHRadiativeEff - High & 1.11         & 0.91         & $3.9\times10^{-2}$       \\ \hline
BHRadiativeEff - Low  & 1.12         & 0.92         & $3.8\times10^{-2}$       \\ \hline
BHJetTvirVel - High   & 1.09         & 0.91         & $3.8\times10^{-2}$       \\ \hline
BHJetTvirVel - Low    & 1.12         & 0.92         & $3.9\times10^{-2}$       \\ \hline
BHJetMassThr - High   & 1.20         & 0.91         & $3.9\times10^{-2}$       \\ \hline
BHJetMassThr - MedHigh    & 1.17         & 0.92         & $3.8\times10^{-2}$   \\ \hline
BHJetMassThr - Low    & 1.03         & 0.93         & $1.8\times10^{-2}$       \\    
\end{tabular}%
\end{table}    

    \subsection{Considering degeneracies with the assumed ionizing background}

    An important factor to consider when analyzing the impact of baryonic physics like feedback on the \lya~forest in simulations is potential degeneracies with the assumed ionizing background or UVB.
    The UVB influences both the temperature and ionization fraction of hydrogen and helium in diffuse gas such as the IGM.
    However, UVB models have a degree of uncertainty and different models can vary by a factor of nearly two at $z=0$ \citep[e.g.][]{Faucher-Giguere:2009UVB,Haardt:2012UVB, Puchwein:2019UVB}.
    In order to use the \lya~forest as a constraint for feedback processes, one must consider what impacts are due to AGN feedback versus the UVB model and separate the two.
    In this work, the only variation between the simulations analyzed are in the AGN feedback model so the impacts seen are due to AGN. 
    However, to determine if these impacts are unique to AGN feedback, and can not instead be captured in changes to the ionizing background, the uncertainties due to the assumed UVB must be factored out.
    
    At low redshifts, where the effects of AGN feedback are most prominent ($z<1$), the impact of the UVB can be modeled using the Gunn-Peterson approximation where the optical depth of the \lya~forest flux is inversely proportional to both temperature and the ionization rate \citep[$\tau \propto T^{-0.7} \Gamma_{HI}^{-1}$,][]{Gunn:1965}.
    This allows for the correction of \lya~flux post processing to simulate what the P1D would look like if a different UVB was used, i.e.\ a different value for $\Gamma_{HI}$. 
    In this work we accomplish this by correcting all of the simulations post-processing to match the mean transmitted flux of the fiducial CAMELS-Simba simulation (the black line in each plot).
    This method, in principle, will catch effects due to changing both the photoionizing and photoheating rates of neutral hydrogen which are modeled by the UVB.
    The results of this analysis are presented in Figure \ref{fig:fig_combined}.
    We show results for $z=0.1$ since it is the redshift bin that the corrected P1D shows more than a 10\% effect due to the AGN feedback variations, and only $A_\textnormal{AGN1}$, $A_\textnormal{AGN2}$, and BHJetMassThr show variations greater than 10\%.
    To analyze these results further we extract three values from each P1D ratio: the maximum and minimum value of the P1D ratio ($A_\textnormal{max}$ and $A_\textnormal{min}$) and the $k$ value at which the P1D ratio crosses 1 ($k_0$). 
    The value for $k_0$ corresponds to what physical scale that the mean flux corrected feedback variant P1D diverges from the fiducial prediction and was extracted via interpolation.
    These values are reported in Table \ref{tab:P1Dratiovalues}.

    The greatest impact is seen for increases in the AGN jet speed ($A_\textnormal{AGN2}$) at the largest $k$ values ($k>3\times10^{-2}$ s/km) as seen through its $A_\textnormal{max}$ value.
    Other parameters that show a maximum change of $\sim20$\% to the P1D is $A_{AGN1}$ for decreases in the value and BHJetMassThr for increases in the value.
    All other parameter variations show a maximum change to the P1D closer to $\sim$10\%.
    For each parameter, $k_0$ lies around the same value (between 0.02 and 0.04 s/km) except for $A_{AGN1}$ high and BHJetMassThr low. 
    For those two variations, the P1D ratio is noisy around the region of interest causing the P1D ratio to cross 1 early. 
    Noise in the ratios is due to the finite sampling of points for $k$.
    The $k$ value 0.04 s/km is roughly around the filtering scale which is the scale at which pressure smoothing dominates and absorbers in the \lya~forest will be dominated by temperature effects like thermal broadening \citep{Gnedin+Hui:1998,Tillman:2025JATIS}.
    This highlights that the impact of AGN feedback can manifest in different ways at the low-$k$ vs.\ the high-$k$ end of the P1D due to the unique role that temperature plays on different scales.
    
    For example, looking at the impact of $A_{AGN2}$ in Figure~\ref{fig:jet_speed}, if the AGN jet speed is increased then the heating due to said jets can reach further into the IGM, ionizing the neutral hydrogen there, and thus suppressing the entire P1D. 
    Simultaneously, this additional heating increases the temperature of the remaining \lya~absorbers, broadening their absorption lines, and causing a further suppression of the high-$k$ end of the P1D.
    This unique effect at the high-$k$ end of the P1D, due to AGN heating, is therefore difficult to capture in the assumed ionizing background which has primarily a normalization effect on the P1D.
    Scaling the strength of the UVB up or down in this case would not capture these effects due to the stronger impacts seen at the high-$k$ end of the P1D.
    
    AGN jets in particular have been found to play an important role not just in this work but in other works exploring \lya~forest statistics in CAMELS and Simba \citep{Tillman:2023AJ,Dong:2024,Tillman:2025} since AGN jets reach distances far beyond the host halo.
    In a broader context, AGN jets in Simba have been found to play a large role in redistributing baryons in the simulation \citep{Delgado:2023, Khrykin:2024, Wright:2024,Sorini:2024,Medlock2025, Gebhardt2026}.
    This highlights that while isotropic feedback effects on the IGM in simulations may be modeled through the assumed metagalactic ionizing background, observationally motivated AGN jets in simulations are more complex.
    This implies that AGN jets require specific consideration as a sub-grid model to ensure their effects on the gas beyond the host galaxy can be studied.

    \subsection{Comparison with observational data}

    As previously stated, we refrain from making direct comparisons to observations in this work due to the small box size of the simulations explored herein.
    Furthermore, the most extreme parameter values in the CAMELS simulations typically result in nonphysical results such as a mismatch between the predicted and observed stellar mass function. 
    The benefit in analyzing these simulations lies in the ability to explore the extent to which a parameter may impact a predicted observable.
    Despite this we can still consider which parameter variations may be observable given current observational data.
    Current observations of the \lya~forest exists up to about $z=0.4$, and these observations have errors averaging around 20\% when looking at P1D ratios \citep{Danforth:2016, Khaire:2019}.
    In Figure \ref{fig:fig_combined} we see that only $A_{AGN1}$, $A_{AGN2}$, and BHJetMassThr impact the P1D more than 20\% after the mean flux re-scaling.
    In theory these changes are right on the limit of what is observable given current \lya~forest data.
    Further discrimination between different feedback parameters requires higher precision \lya~forest data.
    The level of precision required may be achieved given a high resolution FUV spectrograph aboard the Habitable Worlds Observatory \citep{2025JATIS..11d2219T}.

\section{Conclusion}\label{s:Conclusion}

    This work explores simulations from the CAMELS project analyzing variations of five different AGN feedback parameters at $z=0.1$, $0.4$, $0.7$, $1.0$, $1.25$, and $2.0$.
    The benefit of these single-parameter variations is through the individual study of the impact of these components of the AGN feedback model on the \lya~forest. 
    The five parameters explored are AGN momentum flux, AGN jet speed, SMBH radiative efficiency, jet velocity threshold, and the SMBH mass threshold for jet feedback. 
    We calculate the predicted P1D for the minimum, fiducial, and maximum values for each of the parameters.
    From the results, we draw the following conclusions.

\begin{itemize}
    \item The speed of the AGN jet, and thus the distance the jets reach, has the greatest impact of the AGN feedback parameters that we consider on the 1D flux power spectrum of the \lya-forest (Figure \ref{fig:jet_speed}).
    \item Similarly, changing the SMBH mass threshold at which AGN jet feedback occurs has a large impact on the \lya~forest (Figure \ref{fig:1P_28}). 
    The P1D is strongly affected by the mass threshold as long as it is greater than $10^7 M_\odot$, which implies that SMBHs with that mass or greater are the primary contributors to the impact seen on the \lya~forest through jet feedback.
    \item Reducing the AGN jet-mode momentum flux increases the amplitude of the low redshift P1D. However, increasing the jet-mode momentum flux does not further decrease the P1D, implying that the fiducial Simba AGN feedback momentum parameter maximizes the effect on the \lya~forest (Figure \ref{fig:flux}).
    \item Heating the jets at injection helps remove neutral hydrogen from the \lya~forest up to a point after which the extra heating actually suppresses further jet feedback by suppressing SMBH growth, resulting in a lower overall impact (Figure~\ref{fig:BHstats}).
    The impact of AGN jets in the Simba simulations cannot be accounted for in the assumed UVB due to the unique effect that jet heating has on the \lya~forest at the high-$k$ end of the P1D. Thus to effectively study the impact that AGN feedback has on the IGM in simulations, the sub-grid model must be constructed with careful consideration on how AGN jets propagate energy and matter to scales beyond the host galaxy.
\end{itemize}

    More focus must be placed on how AGN feedback models are calibrated to fully understand the extent to which they can impact their surroundings and how these impacts fit into the bigger picture of the baryon cycle. 
    AGN feedback models are often calibrated to match the star formation histories, masses, and colors of galaxies, but the inclusion of larger-scale statistics in these constraints can improve feedback models further.
    To further explore the impact that AGN feedback variations have on the \lya~forest, additional statistics may be explored in addition to the P1D.
    These may include the $b$-value distribution, a statistic that complements that of the high-$k$ end of the P1D, the column density distribution, and the flux PDF.
    Additionally, a more thorough analysis of the thermal history in these simulations by analyzing the temperature-density evolution of the IGM may provide a clearer picture of which effects can be uniquely captured by AGN feedback models as opposed to being included in other models, such as the assumed ionizing background.
    This type of analysis can be conducted with a new set of CAMELS IllustrisTNG simulations that vary both galactic feedback parameters and parameters controlling the assumed UVB.
    Additionally, these simulations are run in a large box which will help minimize uncertainties due to host halo mass sampling.

\begin{acknowledgments}
B.B. acknowledges support from NSF grant AST-2407877 . B.B and S.B. acknowledge NASA grant No. 80NSSC20K0500.
B.B. is grateful for the generous support by the David and Lucile Packard Foundation and the Alfred P. Sloan Foundation. 
B.B. thanks the Center for Computational Astrophysics (CCA) of the Flatiron Institute and the Mathematics and Physical Sciences (MPS) division of the Simons Foundation for support.  The Flatiron Institute is supported by the Simons Foundation.
\end{acknowledgments}

\bibliography{mybib}{}
\bibliographystyle{aasjournal}

\end{document}